\documentclass[%
 reprint,
 amsmath,amssymb,
 aps,
]{revtex4-2}

\usepackage{graphicx}
\usepackage{dcolumn}
\usepackage{bm}
\usepackage{gensymb}

\begin{document}

\preprint{APS/123-QED}

\title{Modular chip-integrated photonic control of artificial atoms in diamond nanostructures}

\author{Kevin J. Palm$^{1,\dag,*}$, Mark Dong$^{1,2,\dag,*}$, D. Andrew Golter$^{1}$, Genevieve Clark$^{1,2}$, Matthew Zimmermann$^{1}$,  Kevin C. Chen$^{2}$, Linsen Li$^{2}$, Adrian Menssen$^{2}$, Andrew J. Leenheer$^{3}$, Daniel Dominguez$^{3}$, Gerald Gilbert$^{4,*}$, Matt Eichenfield$^{3,5,*}$, and Dirk Englund$^{2,6}$}
\email{kpalm@mitre.org}
\email{mdong@mitre.org}
\email{ggilbert@mitre.org}
\email{eichenfield@arizona.edu}
\email{englund@mit.edu}

\affiliation {\it $^1$The MITRE Corporation, 202 Burlington Road, Bedford, Massachusetts 01730, USA}
\affiliation{\it $^2$Research Laboratory of Electronics, Massachusetts Institute of Technology, Cambridge, Massachusetts 02139, USA}
\affiliation{$^3$Sandia National Laboratories, P.O. Box 5800 Albuquerque, New Mexico, 87185, USA}
\affiliation{$^4$The MITRE Corporation, 200 Forrestal Road, Princeton, New Jersey 08540, USA}
\affiliation{$^5$College of Optical Sciences, University of Arizona, Tucson, Arizona 85719, USA}
\affiliation{$^6$Brookhaven National Laboratory, 98 Rochester St, Upton, New York 11973, USA}
\affiliation{$\dag$These authors contributed equally}

\date{\today}

\begin{abstract}
A central goal in creating long-distance quantum networks and distributed quantum computing is the development of interconnected and individually controlled qubit nodes. Atom-like emitters in diamond have emerged as a leading system for optically networked quantum memories, motivating the development of visible-spectrum, multi-channel photonic integrated circuit (PIC) systems for scalable atom control. However, it has remained an open challenge to realize optical programmability with a qubit layer that can achieve high optical detection probability over many optical channels. Here, we address this problem by introducing a modular architecture of piezoelectrically-actuated atom-control PICs (APICs) and artificial atoms embedded in diamond nanostructures designed for high-efficiency free-space collection. The high-speed 4-channel APIC is based on a splitting tree mesh with triple-phase shifter Mach-Zehnder interferometers. This design simultaneously achieves optically broadband operation at visible wavelengths, high-fidelity switching ($> 40$ dB) at low voltages, sub-$\mu$s modulation timescales ($> 30$ MHz), and minimal channel-to-channel crosstalk for repeatable optical pulse carving. Via a reconfigurable free-space interconnect, we use the APIC to address single silicon vacancy color centers in individual diamond waveguides with inverse tapered couplers, achieving efficient single photon detection probabilities (15$\%$) and second-order autocorrelation measurements $g^{(2)}(0) < 0.14$ for all channels. The modularity of this distributed APIC - quantum memory system simplifies the quantum control problem, potentially enabling further scaling to 1000s of channels.
\\
\\
Approved for Public Release; Distribution Unlimited. Public Release Case Number 22-4195\\
$\copyright$ 2022 The MITRE Corporation. All rights reserved.
\end{abstract}

\maketitle

\section{Introduction}
Solid-state artificial atoms \cite{Atature2018-ir}, many of which have long-lived quantum memories \cite{Bar-Gill2013-np, Bradley2019-gk, Dutt2007-lj, Stas2022-uv}, can achieve photon-mediated remote-entanglement \cite{Humphreys2018-cd, Hensen2015-lm}, and can be heterogeneously integrated with photonics \cite{Sipahigil2016-om, Wan2020-wi}, are a promising platform for the construction of large-scale quantum networks \cite{Nielsen2010-nz, Wehner2018-ro, Kimble2008-ac}. The networking of these atom-like emitters requires an efficient and high-fidelity optical interface for both reconfigurable optical addressing and collection of photoluminescence (PL) at visible wavelengths. The optical control layer thus presents two challenges: i) scalable high-fidelity manipulation of optical fields at high speeds, which necessitates high-quality optical switches in atom-control photonic integrated circuit (APIC) \cite{Bogaerts2020-bc} platforms and ii) scalable high-efficiency photon collection from remotely addressable single emitters. While previously demonstrated visible-wavelength APIC platforms such as thin-film lithium niobate \cite{Christen2022-la, Desiatov2019-rz, Li2022-ga}, thermally-tuned silicon nitride \cite{Liang2021-ol, Mohanty2020-jp, Yong2022-tj}, and piezoelectrically-actuated silicon nitride \cite{Menssen2022-ss, Dong2021-op, Dong2022-my, Wang2022-pm} all have promise for scalability, none currently combine optically broadband operation, high switching contrast ($> 40$ dB) at nanosecond time scales, and low voltage operation. On the photon collection side, efficient collection has been demonstrated using standard confocal microscopy \cite{Becker2018-mb, Becker2016-mb, Neu2012-td}, by leveraging photonic nanostructures such as immersion lenses \cite{Robledo2011-sl, Hepp2014-vy, Schroder2011-gz} and cavities \cite{Sipahigil2016-om, Riedrich-Moller2014-qj, Benedikter2017-wk, Lee2012-ur, Zhang2018-xo}, or single-channel fiber collection from tapered waveguides \cite{Sipahigil2016-om, Arjona_Martinez2022-is, Bhaskar2017-om}. Collection through a heterogeneously-integrated photonic chip \cite{Wan2020-wi, Andrew_Golter2022-zr, Mouradian2015-wj} at the cost of some optical loss due to the diamond-chip interface has also been reported. To date, these past works treated each side of the optical control layer separately, but there remains an open question of how to combine the requirements of i) and ii) into a single scalable system.

\begin{figure*}
    \centering
    \includegraphics[width=\textwidth]{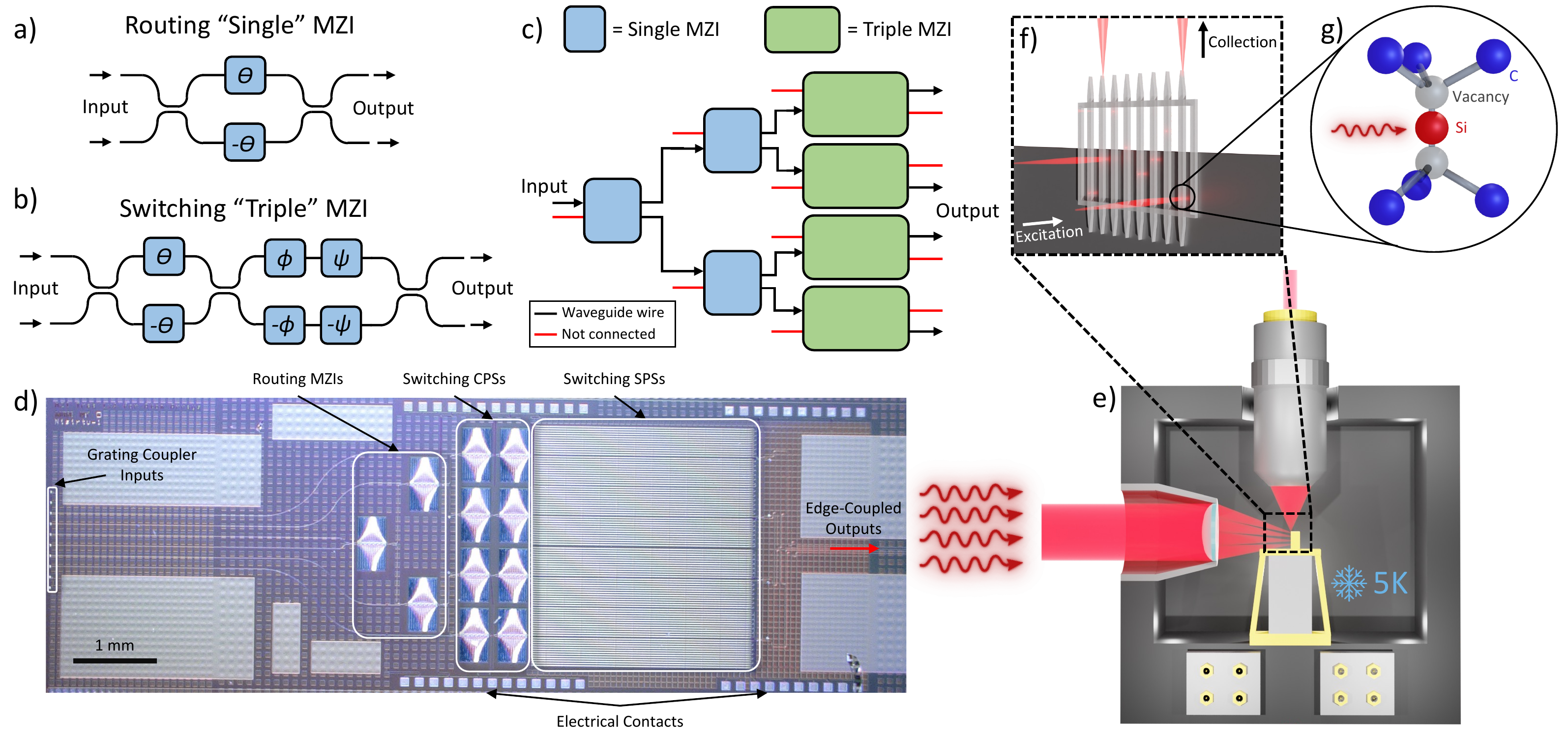}
    \caption{Photonic integrated network switch architecture for local addressing of multiple quantum emitters. a) Routing “single” MZIs to split the single input into each of the four ports and b) switching “triple” MZIs that enable fast arbitrary pulsing of light with high extinction. The routing MZIs consist of a single cantilever phase shifter (CPS) and two 50:50 directional couplers, while the switching MZIs consist of two CPSs, a strain-optic phase shifter (SPS), and three 50:50 couplers. c) Schematic of the binary tree switch design. d) Microscope image of the fabricated integrated network switch with the CPSs and SPSs labeled. Light is input through grating couplers on the left side and collected through edge-coupled outputs on the right. e) Cryostat setup housing the quantum emitters with light from the switch projected through free-space for quantum control experiments. f) Diamond quantum microchiplets with g) implanted Si vacancy color centers with light pulses from the chip controlling the optical emission. The diamond nanostructure allows for high-efficiency collection of the emitter’s emission.}
    \label{fig1}
\end{figure*}

Here we introduce an architecture for the optical control layer consisting of modular piezoelectrically-actuated APICs and diamond microchiplets with implanted single emitters. In this configuration, the excitation and collection optical paths are perpendicular, enabling the inverse tapered diamond waveguides to take advantage of free-space modal conversion for efficient collection through the optical path parallel to the waveguides while maintaining the ability to selectively address a large area of distinct emitters through the perpendicular path. We demonstrate our control scheme by first satisfying requirement i) through our APIC switch, implemented as a 4-channel binary tree mesh \cite{Bogaerts2020-bc} with visible-wavelength switching and power routing capabilities. The APIC’s switching circuit uses an optically-broadband triple-phase shifter design that takes advantage of hardware error correction \cite{Miller2015-ny, Wang2020-wd} and a stronger strain-optic response than previous designs, enabling low switching voltages while maintaining high-contrast ($> 40$ dB) and high-speed ($> 30$ MHz) switching performance. The switch shows negligible cross-talk between channels and enables repeatable arbitrary pulse carving on all four outputs, combined with $> 1$ MHz power balancing between ports. We further demonstrate requirement ii) by applying the APIC to a local group of quantum emitters by projecting the optical output channels onto ion-implanted silicon vacancy color centers (SiVs) \cite{Hepp2014-vy,Sukachev2017-hc} in diamond microchiplets \cite{Wan2020-wi} mounted in a 5K cryostat. Through PL excitation (PLE) and second-order autocorrelation measurements, we demonstrate optical addressing with independent temporal control of four spatially distinct color centers and achieve high (15$\%$) collection efficiency, single emitter linewidths of 152 MHz -  287 MHz, and $g^{(2)}$(0)  of 0.06 - 0.14. The modularity of this architecture allows for easy switching between different sets of quantum emitters by adding different sets of diamond microchiplets into the cryostat setup. Our APIC excitation and diamond collection techniques should enable scalable quantum control of emitters as part of a larger network of quantum nodes.

\begin{figure*}
    \centering
    \includegraphics[width=\textwidth]{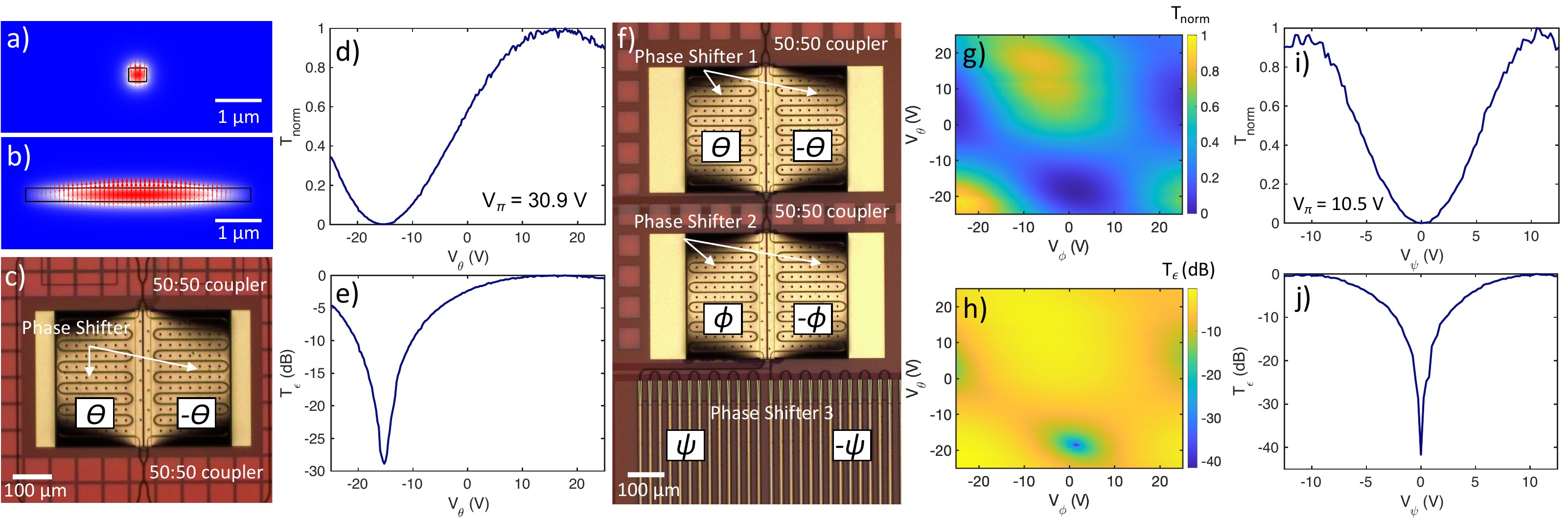}
    \caption{Device performance and calibration. Simulated TM optical waveguide mode for the a) 400 nm waveguides and b) the 5 $\mu$m waveguides in the SPSs. c) Microscope image of a routing MZI. A voltage is anti-symmetrically applied to each side of the phase shifter to give the maximum actuation range. d) Normalized transmission ($T_{norm} = T/T_{max})$ and e) extinction ($T_\epsilon = 10\times log(T_{norm}$)) measured from a single output with the applied voltage to the cantilever swept from -25 to 25 V. A single phase shifter achieves 25-30 dB extinction. f) Microscope image of a switching MZI with three phase shifters. The first two CPSs are calibrated with the SPS held at 0 V to maximize output port extinction. g) Normalized transmission $T_{norm}$ and h) extinction $T_\epsilon$ plots measured from sweeping the applied voltages of the two CPSs. The addition of the second cantilever compared to the single MZI allows for the output extinction to exceed 40 dB. i) Normalized transmission $T_{norm}$ and j) extinction $T_\epsilon$ plots for the SPS, calibrated after the two CPSs in the switching MZI.}
    \label{fig2}
\end{figure*}

\section{Photonic integrated switch design and operation}
The schematic of our APIC-to-diamond control architecture is as follows. The APIC design consists of a “single” routing Mach-Zender Interferometer (MZI) (Fig. \ref{fig1}a) and a “triple” switching MZI (Fig. \ref{fig1}b) arranged in a binary tree architecture (Fig. \ref{fig1}c). A single cantilever phase shifter (CPS) \cite{Dong2022-my} in the routing MZIs directs the desired amount of light to the appropriate outputs. The switching MZI uses three phase shifters: two CPSs that enable optically broadband and high-fidelity routing ($> 40$ dB) for cross and bar ports using hardware error correction robust to fabrication imperfections \cite{Wang2020-wd} and a third, strain-optic phase shifter (SPS) \cite{Stanfield2019-lc, Dong2021-op}, that enables a fast phase response for on-off switching of the output channel. During operation, a CPU controller programs the two CPSs to route the light to a dump port while the SPS is held at 0 V. We then can send an arbitrary pulse sequence to the SPS to switch the light to the output port without having to change the applied DC voltages to the CPSs.

A microscope image of the APIC is shown in Fig. \ref{fig1}d, with the different phase shifters and electrical contacts labeled. We input light into the chip with an optical fiber array through a single grating leading to the routing MZIs, while other inputs are only used for device calibration. We then collect the edge-coupled light from each output with a high-NA objective, enabling imaging of the outputs into any system for optical control experiments. Figure \ref{fig1}e shows the optical imaging schematic where the output channels are projected into a cryostat to use for optical control of quantum emitters in diamond waveguides (Fig. \ref{fig1}f), such as SiVs (Fig. \ref{fig1}g). This configuration enables perpendicular excitation of the diamond waveguides, with the single photon fluorescence from the emitters coupling to the diamond waveguide mode and emitting vertically for collection through inverse tapered couplers, as shown in Figure 1f. This free-space collection allows for efficient and scalable detection due to low-loss collection optics that are robust to misalignment when compared with fiber coupling or PIC integration. Electrical control of the integrated optical components is made through a custom printed circuit board (PCB) with wire bonds to the APIC. Commercial arbitrary waveform generator boards, embedded in a National Instruments PXIe system, control the CPSs and SPSs. A single board with 22 active channels controls the CPSs, providing $\pm$ 25 V, and two boards with four channels of arbitrary waveform generation each control the SPSs, providing $\pm$ 2.5 V. High-speed amplifiers on the PCB amplify the signals to the SPSs to $\pm$ 12.5 V. See Supplementary Sections 1 and 2 for more details on the optical and electrical components of the system.

Figure \ref{fig2} summarizes the APIC characterization and calibration by monitoring the transmission of each edge-coupled optical output. For all optical tests, we use 737 nm wavelength laser light coupled into the TM mode of the on-chip 400 nm wide by 300 nm thick silicon nitride waveguides (modal shape simulated in Fig. \ref{fig2}a), which adiabatically expand to 5 $\mu$m wide in the SPS (Fig. \ref{fig2}b) to increase strain-optic sensitivity \cite{Stanfield2019-lc, Dong2021-op}. The less-confined TM mode takes advantage of a higher photoelastic responsivity when compared to the TE mode \cite{Gyger2020-le}, resulting in a lower $V_\pi$ of the phase shifter than previously reported \cite{Dong2021-op}. Our DC calibration results for the routing MZIs (Fig. \ref{fig2}c) are shown in Fig. \ref{fig2}d-e and for the switching MZIs (Fig. \ref{fig2}f) are shown in Fig. \ref{fig2}g-j, highlighting the low-voltage operation of the SPS for switching and high on-off extinction ratios. These high extinction ratios for the triple-phase shifter are enabled by the second CPS accounting for fabrication imperfections in the 50:50 directional couplers. For calibration data for each of the output ports, see Supplementary Section 3.

\begin{figure*}
    \centering
    \includegraphics[width=\textwidth]{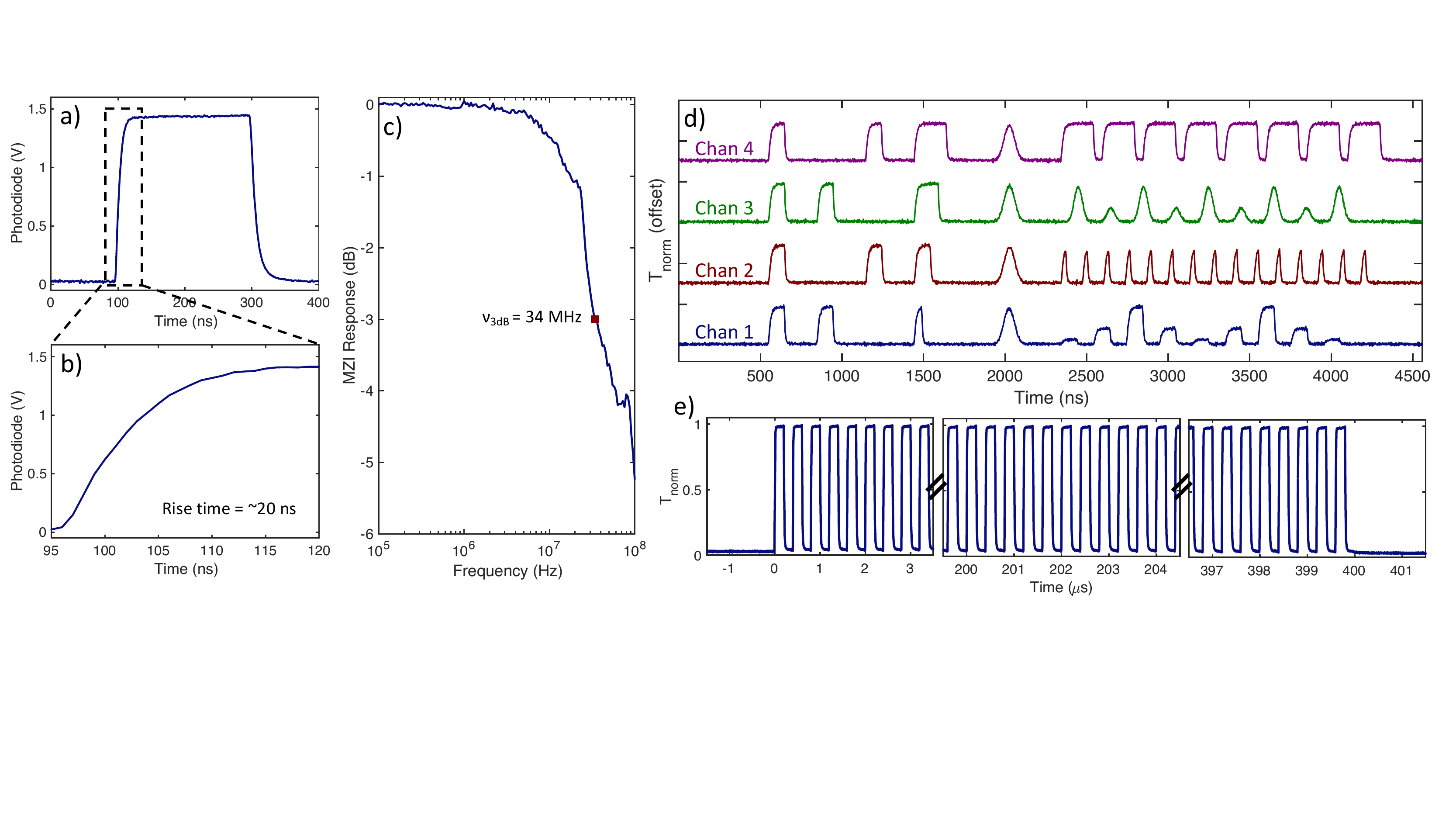}
    \caption{High-speed device pulsing qualification. a) Measured 200 ns pulse from an output port of the chip and b) inset of the pulse showing a $\sim$20 ns rise time from the SPS. c) Normalized modulator response for a 3V sinusoidal signal showing the -3 dB cutoff at $\nu_{3dB} = 34$ MHz. d) Pulsing scheme showing the capabilities of the binary tree for arbitrary pulsing schemes. Each output can be pulsed at arbitrary times, lengths, shapes, and amplitudes. e) Repeated 200 ns pulses with a 50$\%$ duty cycle to measure the consistency of our device. The standard deviation of the integrated pulse area is $6.8 \times 10^{-4}$ for 1000 consecutive pulses.}
    \label{fig3}
\end{figure*}

\section{Pulse Characterization and Stability}
We tested the optical pulse carving of our switch by applying representative pulse sequences to each of the SPSs in the switching MZIs. The “off” state of the output is defined to be 0 V due to the calibration procedure, and the full “on” state is achieved by applying the experimentally determined cross-state voltage. Pulses of varying amplitudes below the maximum are created by setting the applied voltage between these cross and bar states. Using time-resolved measurements on a 125 MHz photodiode, we found rise and fall times of $\sim$20 ns when programming a 200 ns pulse (Fig. \ref{fig3}a,b) for all channels. The small-signal frequency-resolved modular response (Fig. \ref{fig3}c) indicates a -3 dB cutoff at $\nu_{3dB}$ = 34 MHz, allowing for $> 30$ MHz optical control of each channel. The device can also be run at higher modulation speeds ($>100$ MHz) with a trade-off of lower responsivity ($< -6$ dB).

To explore the optical control programmability, we tested various pulse sequences. Figure \ref{fig3}d shows the resulting measurement of each of the outputs and shows four different capabilities of this system: i) Any set of outputs can be pulsed simultaneously, ii) each pulse width can be independently manipulated, iii) the waveform can be temporally amplitude modulated into different shapes, such as square or Gaussian, and iv) the pulse height can be independently set. With these criteria met, our chip has the ability to create a full set of quantum rotations \cite{Li2011-bt}. Furthermore, we measured the consistency of the pulsing of our device by applying repeated 200 ns pulses with 200 ns intervals and measuring the deviations in each pulse. We find a pulse area consistency (1$\sigma$ standard deviation) of $6.8 \times 10^{-4}$ for 1000 pulses, showing robust pulse uniformity. Examples of these pulses from the beginning, middle, and end of this pulse sequence are shown in Fig. \ref{fig3}e. Lastly, we did not observe crosstalk from either thermal, electrical, or piezo effects between the different phase shifters (details in Supplementary Section 4).

\begin{figure*}
    \centering
    \includegraphics[width=\textwidth]{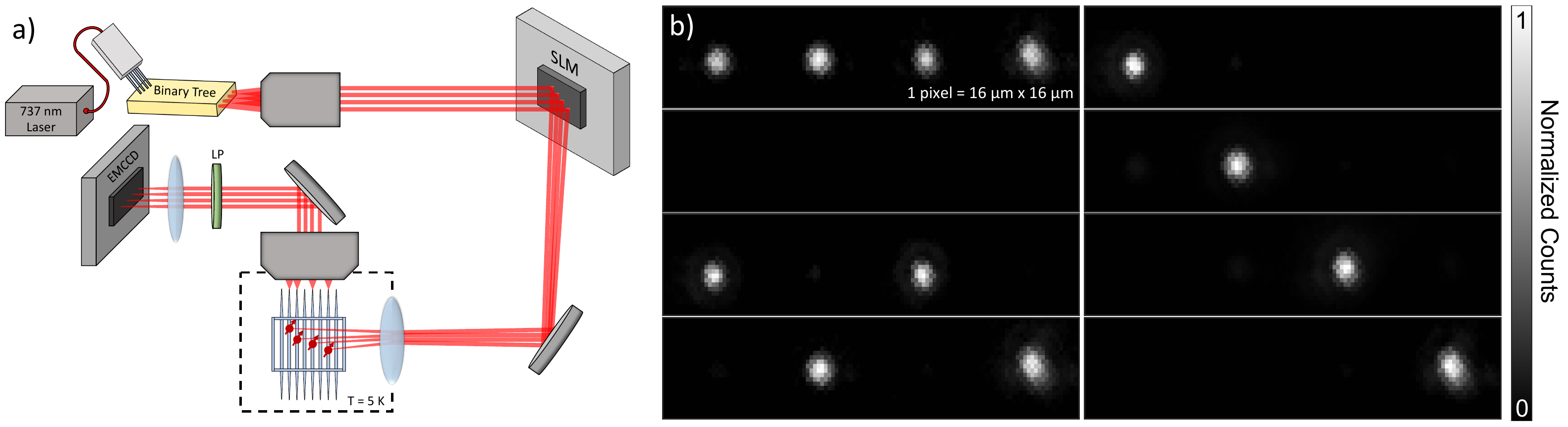}
    \caption{Independent optical control of Si vacancy color centers. a) Experimental setup. 737 nm laser light is input into the binary tree through grating couplers. The four outputs are imaged onto a diamond microchiplet with the use of an SLM to steer the beams onto individual SiVs. The emission of the SiVs is collected, the PSB is filtered out with a 750 nm long pass filter (LP), and imaged onto an EMCCD. b) Simultaneous PLE measurements on four different diamond waveguides. By driving the APIC, emission from emitters in each waveguide can be independently controlled with high extinction. Each panel shows a different iteration of outputs being driven, showing complete independence of emitter emission.
}
    \label{fig4}
\end{figure*}

\begin{figure*}
    \centering
    \includegraphics[width=\textwidth]{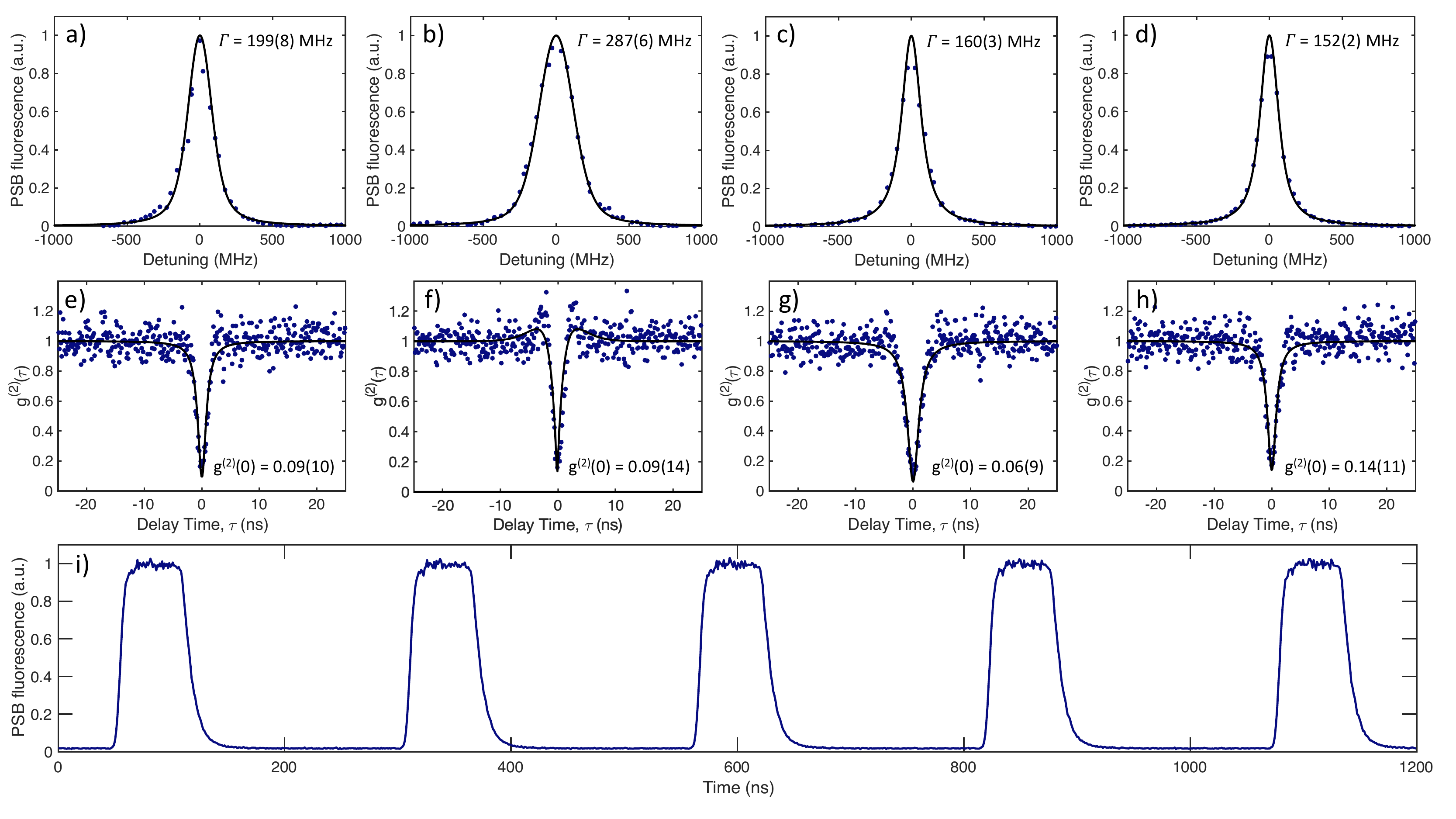}
    \caption{Direct addressing and temporal control of single SiV emitters. a-d) PLE spectrum of single SiVs excited with the APIC. Each vacancy is excited with light from a different APIC channel. e-h) Autocorrelation measurements of the same single SiVs. For each emitter, $g^{(2)}(0) < 0.14$, well below the 0.5 threshold to demonstrate single photon emission. i) Pulsed fluorescence demonstrating temporal control of the emission of a single emitter. Data shown is integrated over 3 min.}
    \label{fig5}
\end{figure*}

\section{Independent addressing of multiple single SiVs}
To demonstrate the applicability of the APIC, we used it to resonantly drive individual emitters within an ensemble of SiVs. As shown in Fig. \ref{fig4}a, the APIC projects each port perpendicularly onto separate diamond waveguides in a cryostat. The diamond waveguides are fabricated with inverse tapered end couplers oriented towards the collection path, allowing for a high collection efficiency of 15$\%$ (see Methods for full diamond fabrication information and Suplementary Section 5 for collection efficiency calculation). The inverse tapers confine the emitted PL to an NA much smaller than that of the collection optics, allowing for scalable collection. In the excitation path, we include a spatial light modulator (SLM) for small spatial adjustments to each projected beam. This allows us to independently steer each excitation spot to specific SiVs in the diamond waveguides. We note that once the SLM is initially programmed, it is kept static over the course of the experiment, making its slow reconfiguration time ($\sim$100 Hz) inconsequential for the excitation experiments. We resonantly excite each of the SiVs while collecting the phonon sideband (PSB) emission using a 750 nm long pass filter to remove excess pump light. We projected this fluorescence onto an electron-multiplying charge-coupled device (EMCCD). Figure \ref{fig4}b shows acquisitions of 30 seconds of the collected fluorescence normalized to the brightest point of each image, with no further image processing. This sequence shows independent and simultaneous optical control of SiVs in four different diamond waveguides. Due to variations in the local strain throughout the diamond, the zero-phonon lines (ZPLs) have an inhomogeneous distribution that exceeds the excitation laser linewidth. To collect SiV emission from multiple waveguides simultaneously, we increased the temperature of the diamond samples to broaden the ZPL linewidths so that they are spectrally overlapping. Thus, for these images, we likely addressed multiple emitters in each diamond waveguide due to the high density of SiVs in our sample ($>50$ emitters per waveguide).

However, to show the applicability of this scheme for controlling individual single emitters, we cooled the diamond sample to a base temperature of 5 K and repeated the excitation scheme with each channel projected onto a spectrally resolved SiV. Figure \ref{fig5}a-d shows the PLE frequency scans for SiVs in four different waveguides, demonstrating linewidths $< 290$ MHz. Second-order correlation measurements indicate strong antibunching, with a normalized $g^{(2)}$(0)  ranging from 0.06 $\pm$ 0.09 to 0.14 $\pm$ 0.11, well below the 0.5 threshold for single photon emission (Fig. \ref{fig5}e-h). We find an average emitter lifetime of 1.76(1) ns (see Supplementary Section 5), consistent with other measurements on ion-implanted SiVs \cite{Wang2005-le}. With the outputs of the MZI tree projected on these emitters simultaneously, we send pulse sequences to temporally control the SiV emission. An example pulse train is shown in Fig. \ref{fig5}i, where we repeatedly pulse one of the channels (Channel 3, Fig. \ref{fig5}c,g) with 100 ns pulses and a period of 250 ns and collect the fluorescence on a time-resolved avalanche photodiode, demonstrating temporal control of a single photon source.

\section{Discussion}
We introduced and demonstrated a scalable optical control system for individual addressing of quantum atom-like emitters. The modularity of the APICs and diamond microchiplets is scalable to 1000s of ports and can be integrated with CMOS control electronics for VSLI devices. Operating voltages can be further reduced by allowing for a trade off of extinction and applied voltage, i.e. if only 30 dB extinction is required then the SPS can be pulsed with $< 2.5$ V applied signal. The diamond collection architecture is also readily scalable, with high efficiency collection of many waveguides enabled by the modal conversion of the waveguides to a 0.26 NA (See Supplementary Section 6). With the collection optics used in this setup, this allows for the scaling to 2975 waveguides with 3 $\mu$m spacing between waveguides in a linear array without a loss of collection efficiency. The losses on the chip currently limit the scalability of the platform, with a total measured insertion loss of -19.2 dB. This loss is dominated by a low grating coupler efficiency of $10\%$, which can be improved with design and fabrication iterations (See Supplementary Section 7 for improved grating coupler results $> 40\%$).

Future work will use this platform for running independent optical control schemes of quantum emitters. Using already demonstrated strain tuning \cite{Meesala2018-xq, Clark2022-zr}, we envision a second chip built from the same APIC platform that allows for spectral matching of quantum emitters, a necessary functionality for quantum computation. More broadly, the broadband \cite{Dong2021-op, Dong2022-my} APIC technology can be applied to other optically trapped atomic systems \cite{Debnath2016-tu, Niffenegger2020-vx, Ebadi2021-ac, Pino2021-ev, Lambrecht2017-fi, Levine2019-nl, Graham2022-ej} and will enable near-future experiments in the area of optical quantum control.

\section{Methods}
\subsection{PIC Calibration}
To begin a calibration, the light from the first output channel is focused onto a photodiode. An iris is used to ensure that only the light from the active output is being measured. A single laptop controls all of the equipment in the experiment and is able to set the applied voltages and query the measured values from the powermeter. We first calibrate the routing MZIs. The applied voltage is swept from -25 V to 25 V in increments of 0.1 V with a power reading at each interval. The voltage is applied differentially to the CPS, with +V being applied to one cantilever and -V applied to the other, nominally giving a $\theta$ and $-\theta$ phase shift for each path respectively. During the calibration, all other phase shifters’ voltages are held constant. To find the cross and bar states, we fit an offset sine curve to the data and take the maximum and minimum values.  

Next, we calibrate the triple-phase shifter switching MZIs. We begin by setting the voltage of the SPS to 0 V, and then calibrate the CPSs. Since the total extinction of this MZI is dependent on the relationship between $\theta$ and $\phi + \psi$, we do a nested two dimensional sweep of the applied voltages from -25 V to 25 V in increments of 0.25 V. The cross and bar states are found by fitting a two dimensional sinusoid to the data and taking the maximum and minimum values respectively. We then set the two CPSs to their bar state (minimum transmission) and calibrate the SPS by sweeping the voltage from -25 V to 25 V in increments of 0.1 V. The SPSs are also operated differentially in a push-pull configuration. We fit an offset sine curve to this data to find its cross and bar states. With how we set up this calibration, the bar state is defined to be 0 V due to the CPSs being set to their bar state. Full calibration results are shown in the Supporting Information Section 3.

\subsection{Diamond Chiplet Fabrication}
For the generation of negatively charged SiV in the diamond, we relieved the strained surface of the diamond plate by removing the top 7 $\mu$m using Ar/Cl$_2$ plasma etching followed by O$_2$ etching. The sample was subsequently implanted with Si$^{29}$ at 190 keV with a dose of $5 \times 10^{10}$ ions/cm$^2$ (Innovion Inc.). It was then annealed in an ultra-high vacuum furnace ($< 10^{-7}$ mbar) at 1200 $\degree$C and cleaned in a boiling tri-acid mixture (1:1:1 nitric acid, sulfuric acid, and perchloric acid at 345 $\degree$C). A 180 nm silicon nitride (Si$_3$N$_4$) was chemical vapor deposited on the diamond, and patterned using electron-beam lithography and CF$_4$ reactive-ion etching (RIE). We isotropically undercut the diamond quantum microchiplet (QMC) using an oxygen inductively coupled plasma (ICP) RIE. Lastly, we submerged the sample in hydrofluoric acid to remove the Si$_3$N$_4$ hard mask and alumina \cite{Mouradian2017-it}.

\subsection{SiV Linewidth and Autocorrelation Measurements}
The diamond sample used in these experiments was fabricated into a QMC \cite{Wan2020-wi} as described above. We then broke the QMC into individual waveguides and placed them overhanging the edge of a cleaved Si chip using tungsten tips. We mounted the Si chip vertically in the Montana cryostat to enable perpendicular excitation and collection.

When measuring the individual SiV emitters, we coupled a single waveguide mode at a time to a multimode fiber for high-efficiency collection. We fit the emitter’s PLE linewidth scans with a Voigt profile using the Nelder-Mead simplex algorithm for the fit optimization. For the autocorrelation measurements, we used a 50:50 fiber splitter to send the light to two APDs in a Hanbury-Brown-Twiss setup. During these measurements, we input pulsed 532 nm light into the waveguide through the collection objective. We input the minimum amount of repump needed to obtain the maximum count rate, providing maximum charge state initialization. We gated the detectors to only collect data when the repump beam is off.

To fit the $g^{(2)}$ values, we used a Lorentzian fit to the data. 

\begin{equation}
g^{(2)}(\tau) = C_1 +C_2 \left( \frac{\frac{1}{2} \Gamma}{\tau + \left( \frac{1}{2} \Gamma \right) ^2} \right)
\end{equation}

\noindent where $\tau$ is the delay time between coincident counts, $C_1$ and $C_2$ are the offset and scaling factors respectively, $\Gamma$ is the full-width half-max of the emitter spectrum. For the one emitter that showed pronounced bunching behavior, we fit the data to a three level system and added in an overall offset and scaling factor to account for the non-ideality of the data due to dark counts and jitter from the APD. 

\begin{equation}
g^{(2)}(\tau) = C_1 + C_2 \left( 1 - (1 + a) e^{-|\tau|/\tau_1} + a e^{-|\tau|/\tau_2} \right)
\end{equation}

\noindent where $\tau$ is the delay time between coincident counts, $C_1$ and $C_2$ are the offset and scaling factors respectively, a is the scaling factor determining the strength of the photon bunching, $\tau_1$ is the antibunching time constant, and $\tau_2$ is the bunching time constant. When compared to the standard Lorentzian fit, we obtained similar values for $g^{(2)}$(0) (0.09 vs 0.07). The error bars reported in the manuscript correspond to one standard deviation in the fit parameters. 

\begin{acknowledgements}
Major funding for this work is provided by MITRE for the Quantum Moonshot Program. D.E. acknowledges partial support from Brookhaven National Laboratory, which is supported by the U.S. Department of Energy, Office of Basic Energy Sciences, under Contract No. DE-SC0012704 and the NSF RAISE TAQS program. M.E. performed this work, in part, with funding from the Center for Integrated Nanotechnologies, an Office of Science User Facility operated for the U.S. Department of Energy Office of Science. M.D. and M.Z. thank MITRE engineers L. Chan, K. Dauphinais, and S. Vergados for their support in building mechanical and electronic components. K.P. and M.D. thank S. Trajtenberg and Y. S. Duan for additional experimental support and C. Li and Y. Hu for helpful conversations and comments.
\end{acknowledgements}

\section*{Author Contributions}
K.J.P. built the experimental setup and performed the device characterization and calibration experiments. K.J.P. and D.A.G., with assistance from G.C. and M.D., built and performed the SiV direct excitation experiments. K.J.P. performed the data analysis. M.Z. designed the electronic control system. M.D., with assistance from A.M., designed the APIC. M.E. and A.J.L., with assistance from D.D., supervised the APIC fabrication. K.C.C. and L.L. fabricated the diamond microchiplets. K.C.C. performed the diamond waveguide simulations. M.D. and D.E. conceived the experiment and device architecture. M.D., G.G., M.E., and D.E. supervised the project. K.J.P. and M.D. wrote the manuscript with input from all authors.  

\section*{Additional Information}
Supplementary information is available for experimental methods related to programming and calibrating the photonic integrated circuit and collection apparatus.

\section*{Competing Interests}
D.E. is a scientific advisor to and holds shares in QuEra Computing.

\section*{Data Availability}
The data that support the plots within this paper are available from the corresponding authors upon reasonable request.

\bibliography{binaryTree}

\end{document}


\preprint{APS/123-QED}

\title{Modular chip-integrated photonic control of artificial atoms in diamond nanostructures: Supplementary Information}

\author{Kevin J. Palm$^{1,\dag,*}$, Mark Dong$^{1,2,\dag,*}$, D. Andrew Golter$^{1}$, Genevieve Clark$^{1,2}$, Matthew Zimmermann$^{1}$,  Kevin C. Chen$^{2}$, Linsen Li$^{2}$, Adrian Menssen$^{2}$, Andrew J. Leenheer$^{3}$, Daniel Dominguez$^{3}$, Gerald Gilbert$^{4,*}$, Matt Eichenfield$^{3,5,*}$, and Dirk Englund$^{2,6}$}

\affiliation {\it $^1$The MITRE Corporation, 202 Burlington Road, Bedford, Massachusetts 01730, USA}
\affiliation{\it $^2$Research Laboratory of Electronics, Massachusetts Institute of Technology, Cambridge, Massachusetts 02139, USA}
\affiliation{$^3$Sandia National Laboratories, P.O. Box 5800 Albuquerque, New Mexico, 87185, USA}
\affiliation{$^4$The MITRE Corporation, 200 Forrestal Road, Princeton, New Jersey 08540, USA}
\affiliation{$^5$College of Optical Sciences, University of Arizona, Tucson, Arizona 85719, USA}
\affiliation{$^6$Brookhaven National Laboratory, 98 Rochester St, Upton, New York 11973, USA}
\affiliation{$\dag$These authors contributed equally}

\date{\today}

\maketitle

\section{Optical Experimental Setup}
\begin{figure*}
    \centering
    \includegraphics[width=\textwidth]{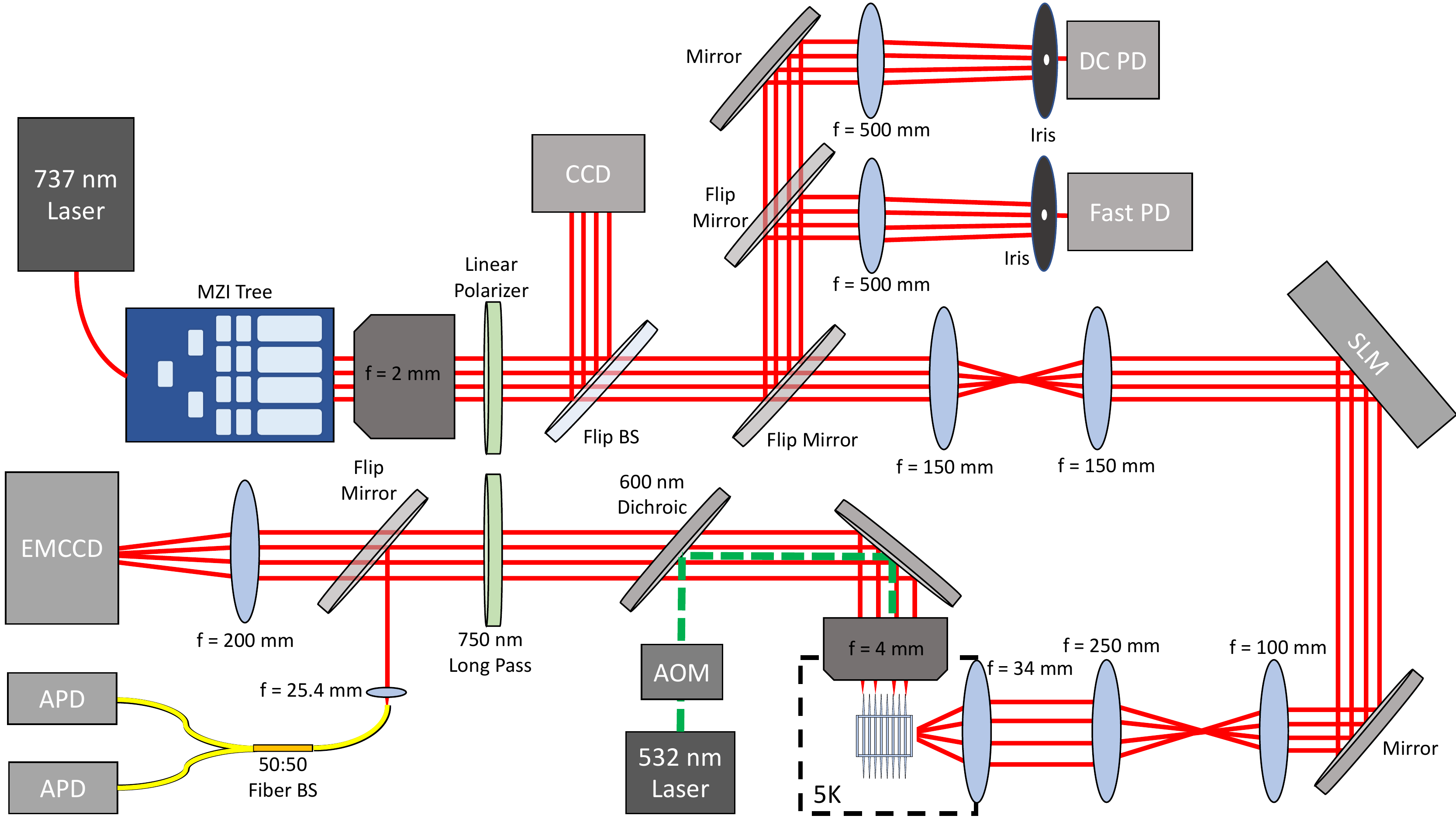}
    \caption{\label{fig:wide}Optical setup for binary tree experiments. Flip mirrors are used to redirect light to different optical detectors for different types of calibration experiments. Labels: f = effective focal length, CCD = charge-coupled device, BS = beamsplitter, SLM = spatial light modulator, AOM = acoustic optical modulator, APD = avalanche photodiode, EMCCD = electron-multiplying charge-coupled device.}
    \label{figS1}
\end{figure*}

Figure \ref{figS1} depicts the optical setup for all of the experiments demonstrated in this manuscript. Laser light was input into the binary MZI tree through a 10-port fiber grating with a tunable laser set at 737 nm (M Squared Lasers). We collected the edge coupled light with a 100x, 0.9 NA infinity corrected objective (Mitutoyo) and filtered the light with a linear polarizer. We then routed the light to different characterization devices using mirrors on flip stages for ease of switching between measurements: 1) a CCD camera for alignment, 2) a DC photodiode (Newport 818-SL) for device extinction characterization, and 3) a 125 MHz fast photodiode (New Focus 1801) for pulsed light characterization. When performing diamond excitation experiments, we routed the light onto a spatial light modulator (SLM) (Thorlabs Exulus) and directed the beams onto vertically-mounted diamond waveguides in a Montana cryostat. The 34 mm lens in the beam path was embedded in the side of the cryostat with a custom inset. A long working distance objective (50x Mitutoyo) collected the emission from the diamond waveguides, whose photons then traveled to the electron-multiplying charge-coupled device (EMCCD), or coupled one of the waveguide modes to a fiber for autocorrelation experiments in a Hanbury-Brown-Twiss setup with two avalanche photodiodes (APD). A 532 nm laser (Obis) was introduced in the collection path and pulsed into the diamond waveguide during autocorrelation experiments to stabilize the SiV charge state \cite{Maity2020-qx}. 

\begin{figure*}
    \centering
    \includegraphics[width=0.8\textwidth]{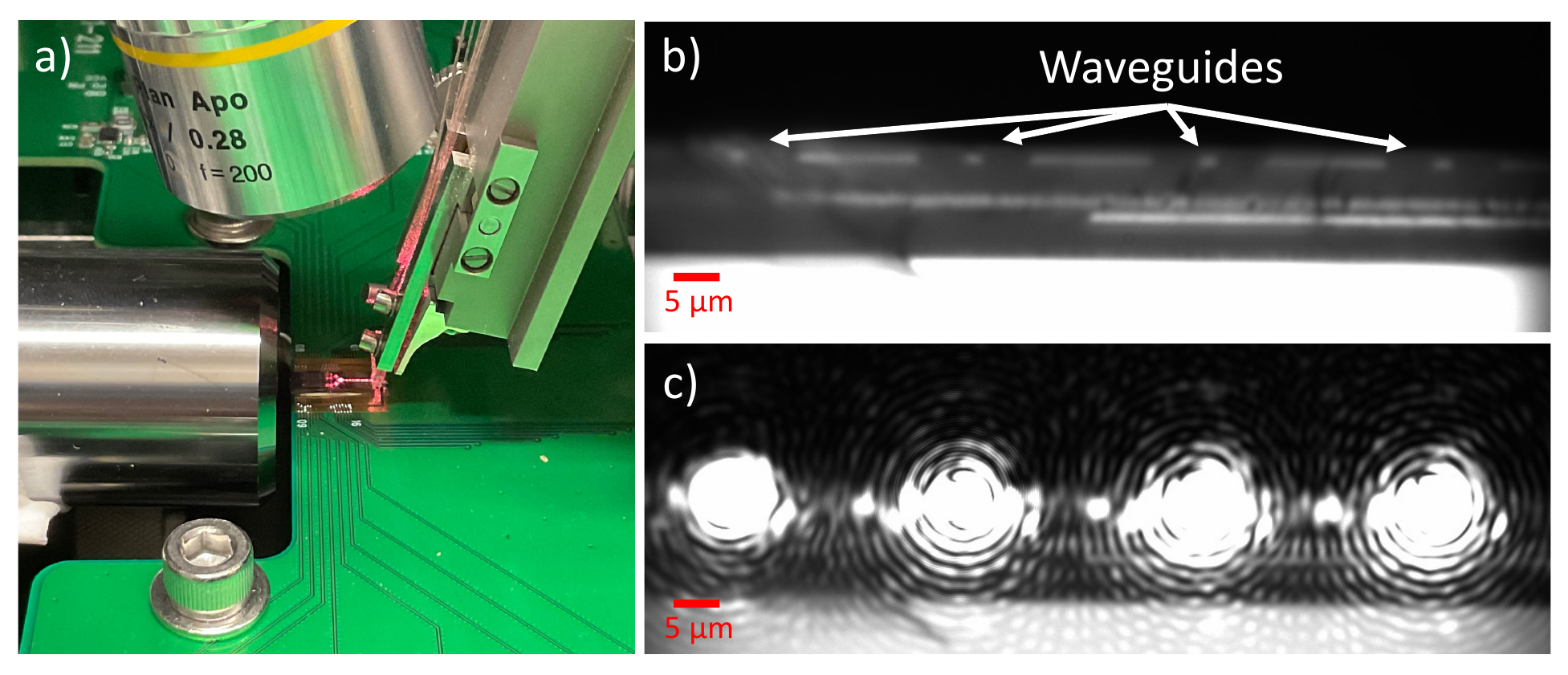}
    \caption{\label{fig:wide}Imaging edge coupled waveguides. a) Picture of the wire bonded APIC onto the PCB. Light is input into the chip through a mounted fiber array (right) into grating couplers. The light is then collected with an objective (left) to be imaged or routed to the cryostat. The top objective is for observing the chip and aligning the fiber array. b) Image of the edge coupled waveguides. c) Same waveguides, but with light being evenly split through each port on the chip.}
    \label{figS2}
\end{figure*}

Figure \ref{figS2}a are images of the APIC coupling setup. The laser light was input through the fiber array at 18$\degree$ (right) and collected with the horizontal objective (left). The vertical objective was used to align the fibers and inspect the chip. Figure \ref{figS2}b shows the imaged edge of the waveguides, with Fig. \ref{figS2}c showing the same waveguides with light being emitted through the chip.

\section{Photonic Integrated Circuit Packaging}
Our photonic chip was fabricated on 200-mm Si technology with a CMOS-compatible fabrication procedure. The binary tree was initially cleaved from the full-wafer die, and then wire-bonded to a custom printed circuit board (PCB). The cantilever phase shifters (CPS) were electrically driven with a 32-channel voltage controller (Marvin Test Systems GX1632e) with a voltage range of $\pm$25 V. The CPSs were driven in a push-pull configuration for maximum phase shift (e.g. if +10 V is applied to the top cantilever, -10 V is applied to the bottom cantilever). The strain-optic phase shifters (SPS) were driven with two 200 MHz arbitrary waveform generator (AWG) PXIe cards (Spectrum M4x.6622), each with four output channels with a voltage range of $\pm$ 2.5 V and 5x amplified on the PCB at a max slew rate of 8000V$\mu$s$^{-1}$ (Texas Instruments THS3491). In the course of any experiment, the SPSs were operated in a push-pull configuration similar to the CPSs.

\section{MZI Calibration Plots}
A labeling of the different phase shifters is shown in Figure S3. For results of a full calibration, see Fig. \ref{figS4} for the routing MZIs, Fig. \ref{figS5} for the CPSs in the triple phase shifters, and Fig. \ref{figS6} for the SPSs.

\newpage

\begin{figure*}
    \centering
    \includegraphics[width=0.8\textwidth]{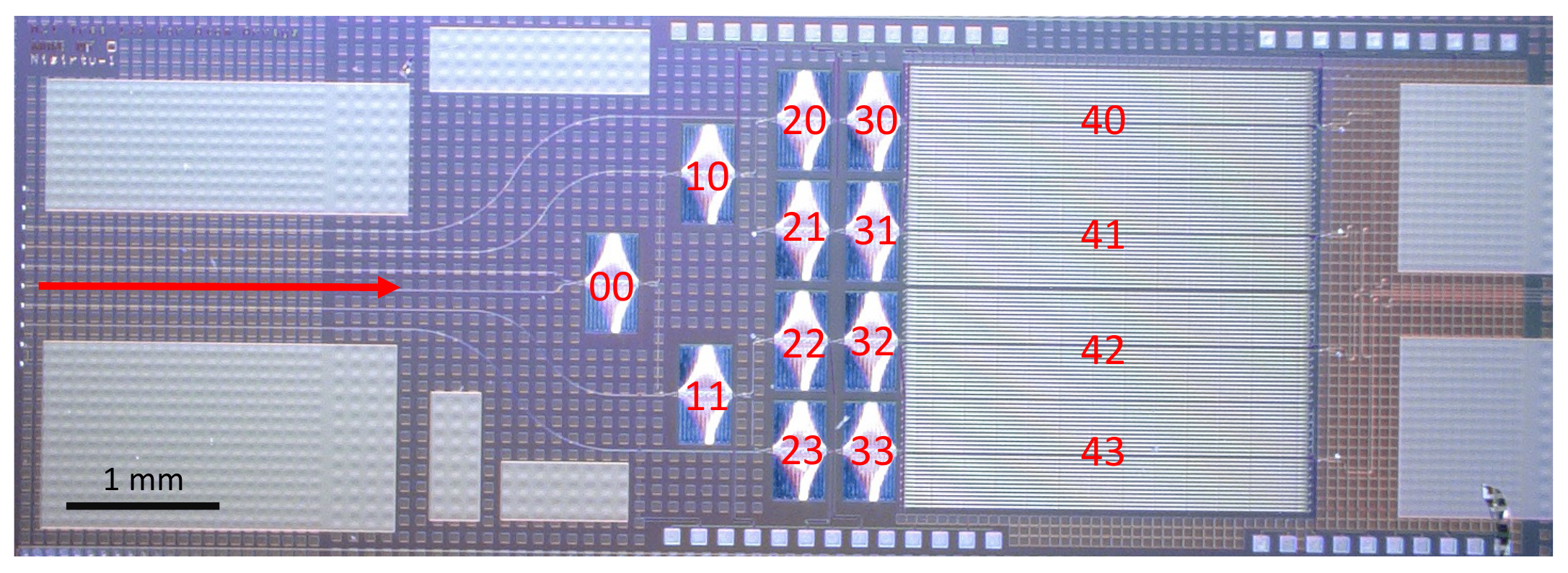}
    \caption{\label{fig:wide}Labeling of different phase shifters for calibration. Light is input from the 6th grating coupler from the top. Output 0 is defined as the top output in this microscope image, with Output 3 as the bottom output.}
    \label{figS3}
\end{figure*}

\begin{figure*}
    \centering
    \includegraphics[width=0.9\textwidth]{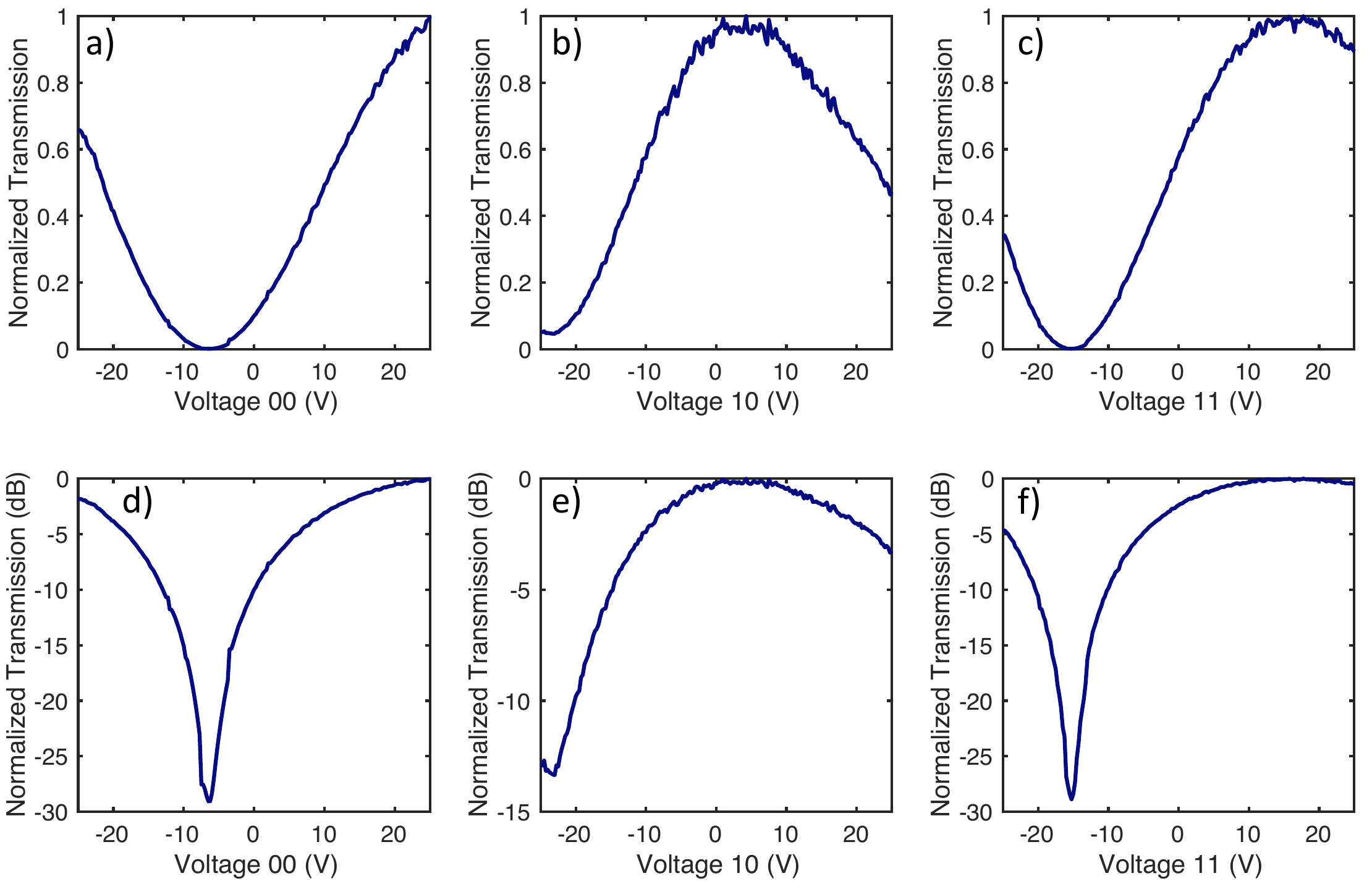}
    \caption{\label{fig:wide}Routing cantilever phase shifter calibration. Voltages are swept from -25 V to 25 V while all other phase shifters are held static. Normalized transmission vs applied voltage for a) PS$_{00}$, b) PS$_{10}$, and c) PS$_{11}$. Each phase shifter has a $V_\pi$ of $\sim$30 V. Extinction measurements for d) PS$_{00}$, e) PS$_{10}$, and f) PS$_{11}$.}
    \label{figS4}
\end{figure*}

\newpage

\begin{figure*}
    \centering
    \includegraphics[width=\textwidth]{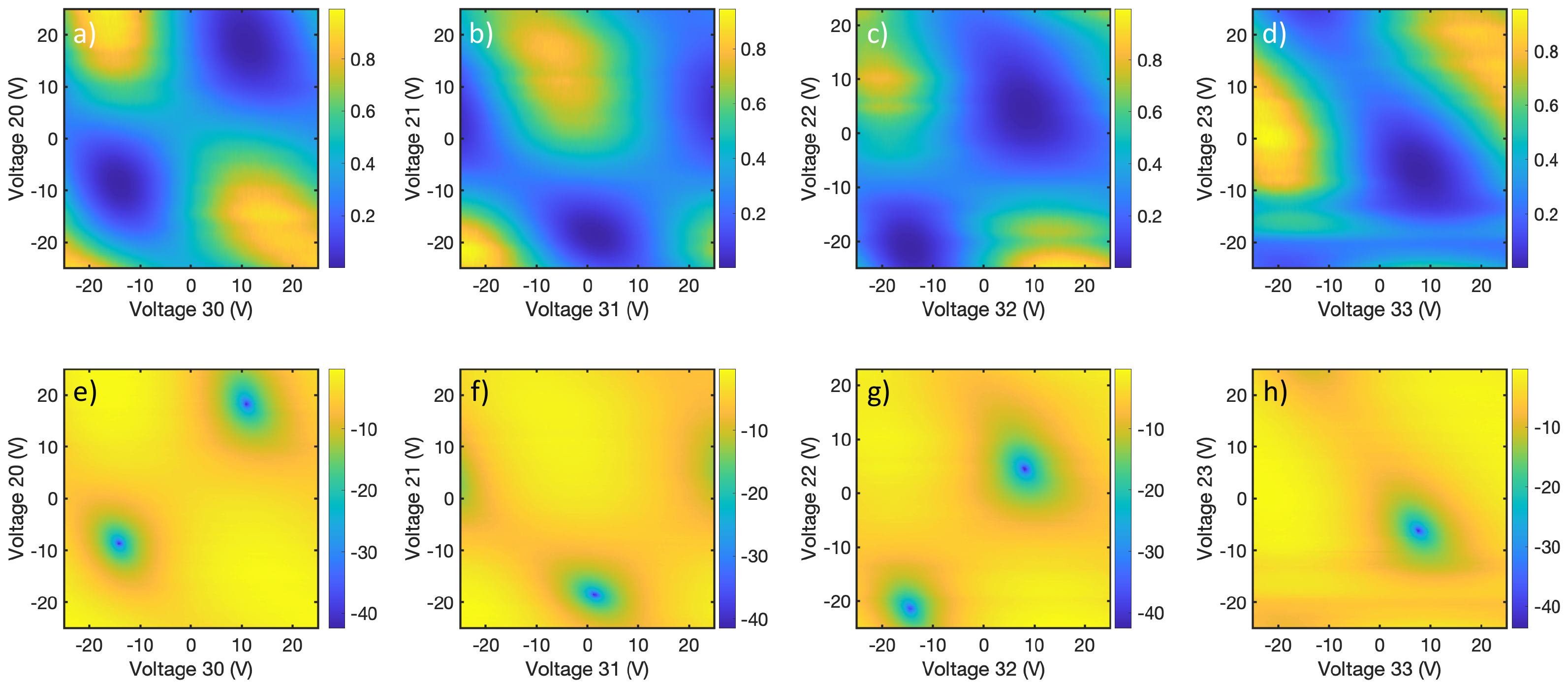}
    \caption{\label{fig:wide} Switching cantilever phase shifter calibration. Voltages are swept from -25 V to 25 V in a nested sweep while all other phase shifters are held static. Normalized transmission vs applied voltage for a) PS$_{20}$ and PS$_{30}$, b) PS$_{21}$ and PS$_{31}$, c) PS$_{22}$ and PS$_{32}$, and d) PS$_{23}$ and PS$_{33}$. Extinction measurements for e) PS$_{20}$ and PS$_{30}$, f) PS$_{21}$ and PS$_{31}$, g) PS$_{22}$ and PS$_{32}$, and h) PS$_{23}$ and PS$_{33}$.}
    \label{figS5}
\end{figure*}

\begin{figure*}
    \centering
    \includegraphics[width=\textwidth]{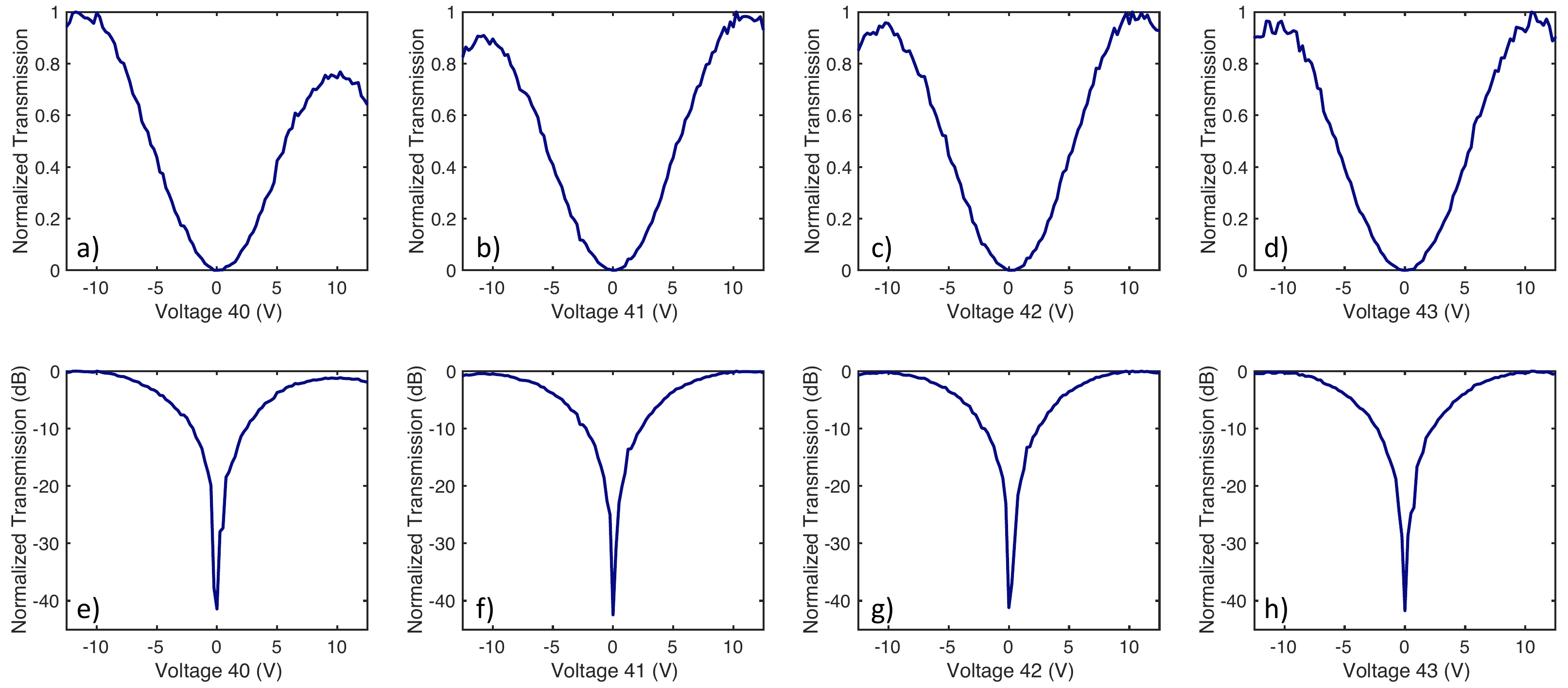}
    \caption{\label{fig:wide}Switching strain-optic phase shifter calibration. Preceding CPSs are set to the bar state (minimum transmission) and then the voltage of the SPS is swept from -25 V to 25 V. Normalized transmission vs applied voltage for a) PS$_{40}$, b) PS$_{41}$, c) PS$_{42}$, and d) PS$_{43}$. Extinction measurements for e) PS$_{40}$, f) PS$_{41}$, g) PS$_{42}$, and h) PS$_{43}$.}
    \label{figS6}
\end{figure*}

\clearpage

\section{Crosstalk Investigation}
\begin{figure*}
    \centering
    \includegraphics[width=0.7\textwidth]{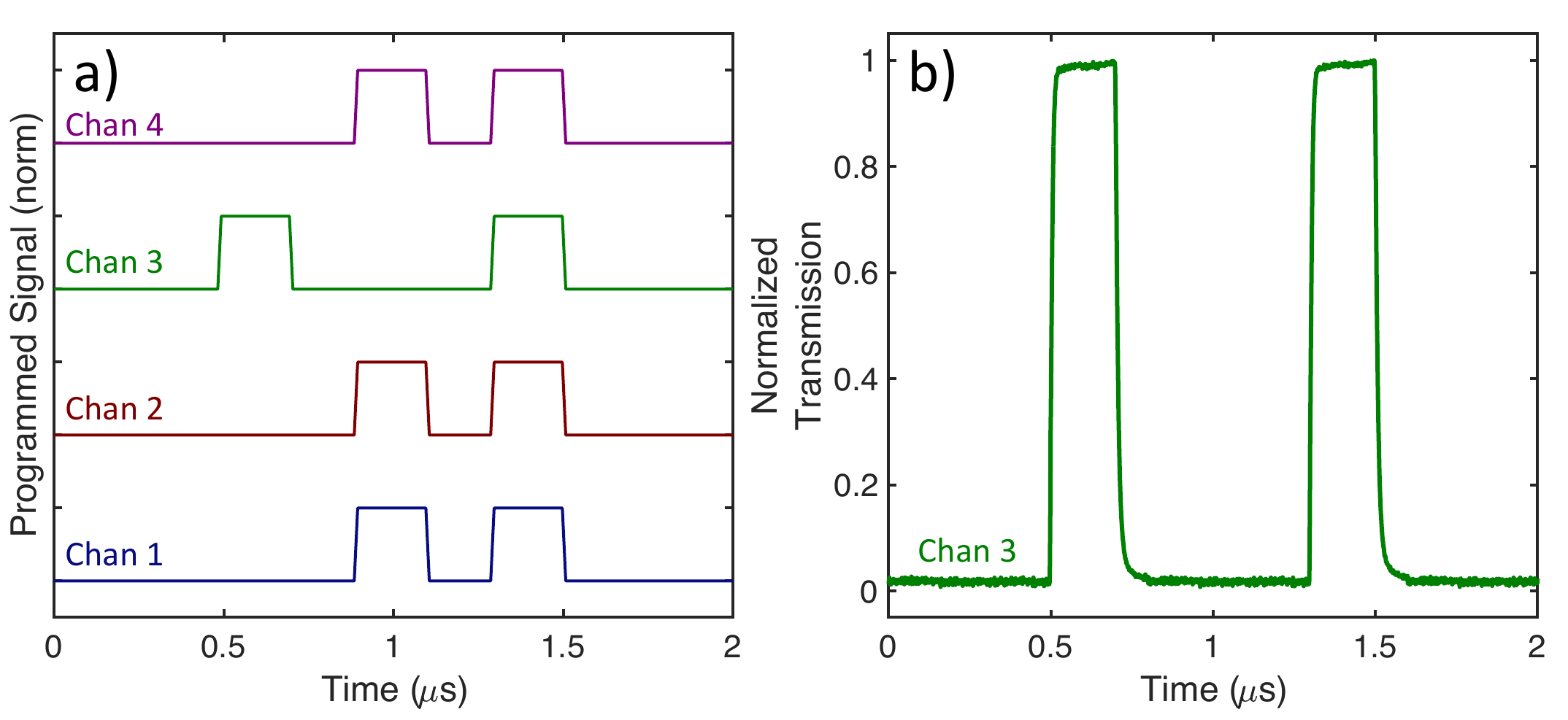}
    \caption{\label{fig:wide}Crosstalk investigation. a) Programmed signal to each of the four SPSs. Channel 3 was imaged onto the power meter for measurement. The first pulse on Channel 3 is a control pulse for comparison, followed by the other three channels pulsing, and ending with all four channels pulsing. b) Measured power of Channel 3. We do not measure any cross talk above the noise floor between any of the channels.}
    \label{figS7}
\end{figure*}

To investigate whether adjacent SPSs had any effect on each other, we ran crosstalk experiments. For these experiments, we set a pulse series to the four SPSs and measured if signals sent to adjacent SPSs had any effect on the measured output. An example measurement is shown in Fig. \ref{figS7}. In this experiment, we had an initial control pulse, a pulse of all other channels except for the measured channel, and ending with a pulse of all four channels. We did not measure differences between the control pulse and all four of the channels pulsing. We also did not measure any output above the noise floor when the other three channels are pulsed and the active channel remains in the bar state. 

\section{Collection efficiency calculation and lifetime measurement}
\begin{figure*}
    \centering
    \includegraphics[width=\textwidth]{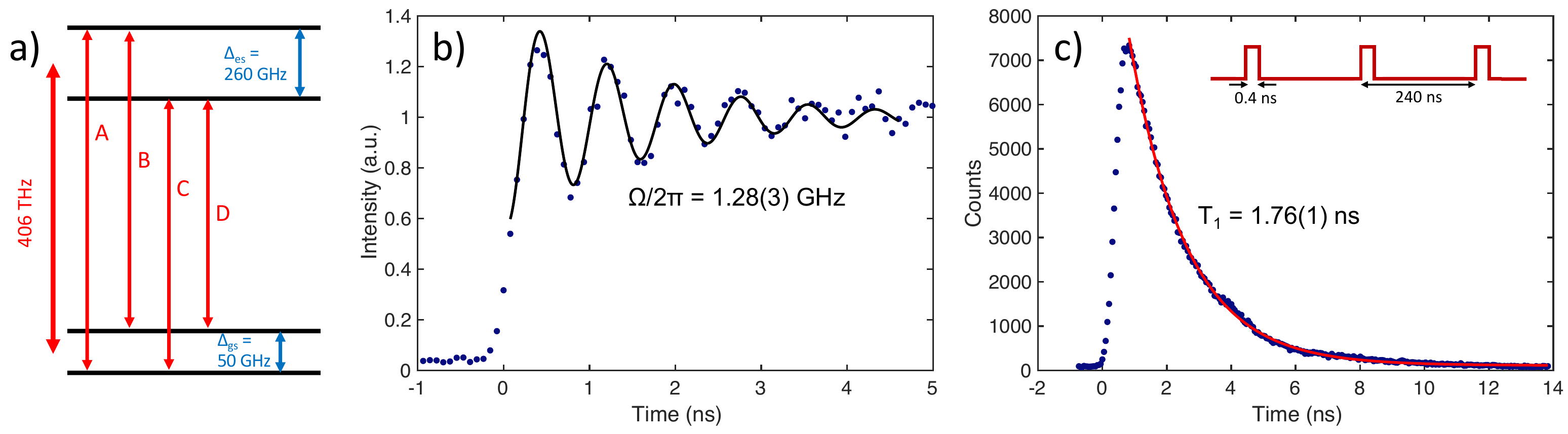}
    \caption{\label{fig:wide}Collection efficiency measurement. a) Electronic structure of the SiV color center. The ground and excited states are split by spin-orbit coupling. For these experiments we drive the C transition. b) Rabi oscillations excited by a 20 ns pulse resonant on the C transition. We fit the oscillations with an exponentially damped sinusoid and find a Rabi frequency of $\Omega/2\pi$ =1.28(3) GHz, corresponding to an optical $\pi$-pulse of 400 ps. c) Pulsed excitation of a single SiV with repeated $\pi$-pulses. Exponential fit of the decay gives a lifetime of $T_1 = 1.76(1)$ ns. Coupling efficiency is found by summing the total counts in the decay and normalizing by applied $\pi$-pulses.}
    \label{figS9}
\end{figure*}

We used a resonant pulsing scheme to measure the collection efficiency of our system. We attached an electro-optic modulator (EOM, iXblue) on the input of our laser source and drove it with an AWG (Tektronix 70001B) to achieve the short excitation pulses necessary for the measurements. For all of these measurements, we drove the C transition of the SiV (Fig. \ref{figS9}a). First, we excited a single emitter with a 20 ns resonant pulse and collected the time-tagged sideband of the emission. We fit the resulting data with an exponentially damped sinusoid to extract the Rabi frequency for our input laser power (Fig. \ref{figS9}b). This fit gave us an optical $\pi$-pulse of 400 ps. We then applied this optical $\pi$-pulse repeatedly, with one $\pi$-pulse every 240 ns to ensure the emitter decayed into the ground state and thermalized, and collected the time-tagged sideband emission. Every 400 $\mu$s we applied a 1-$\mu$s 532 nm repump pulse to initialize the SiV to the negative charge state. Fig. \ref{figS9}c shows the resulting data. We fit the exponential decay of the counts and find a lifetime of $T_1 = 1.76(1)$ ns. 

To find the collection efficiency of our apparatus, we subtracted the background in Fig. \ref{figS9}c, sum the remaining counts, and divided by the total number of applied $\pi$-pulses. This gave us a 0.038$\%$ chance of measuring a photon in the sideband per pulse. A few corrections onto this value must be made to find the accurate collection efficiency. We experimentally measured the transmission percent of the SiV emission through our optical filters by measuring a photoluminescence spectra with 532 nm off-resonant excitation with the 750 nm longpass filter in and out of the collection path, and obtained a sideband transmission of 15$\%$. During this measurement a 600 nm longpass filter was included to block the excess 532 nm pump that coupled to the waveguide mode. In our setup, we did not directly control the electron distribution between the two ground states split by spin-orbit coupling. This population distribution is determined by \cite{Jahnke2015-ct}:

\begin{equation}
\frac{P_{LB}}{P_{UB}} = e^{-h \Delta_{gs}/k_B T}
\end{equation}

\noindent where $P_{LB}$ and $P_{UB}$ are the populations in the lower and upper ground states respectively, $h$ is Planck’s constant, $k_B$ is the Boltzmann constant, $\Delta_{gs}$ is the ground state splitting, and $T$ is the temperature. At 5 K, we find that the electron is in the lower ground state with 62$\%$ probability. We then accounted for an APD measurement efficiency of 55$\%$ at 737 nm (Excelitas Technologies SPCM-AQRH) and the SiV quantum efficiency of 5$\%$ \cite{Neu2012-td, Benedikter2017-wk}. Applying these four corrections, we obtained a final collection efficiency of 15$\%$. We note that this collection efficiency could potentially be higher as the SiV was not initialized to the correct charge state with 100$\%$ fidelity. Continuous wave measurements with 532 nm excitation indicated that the off-resonant repump put the vacancy in the correct charge state with 60$\%$ probability, with no emission recorded when the vacancy was not in the correct charge state. We also note that there is uncertainty in the quantum efficiency value of the emitter, as different nanostructuring schemes cause for different modifications to this value \cite{Riedrich-Moller2014-qj}. Our design could be further improved by adding a Bragg reflector to the back half of our diamond waveguide, which potentially increases our collection efficiency by a factor of two.

\section{Diamond waveguides NA}
\begin{figure*}[b]
    \centering
    \includegraphics[width=0.4\textwidth]{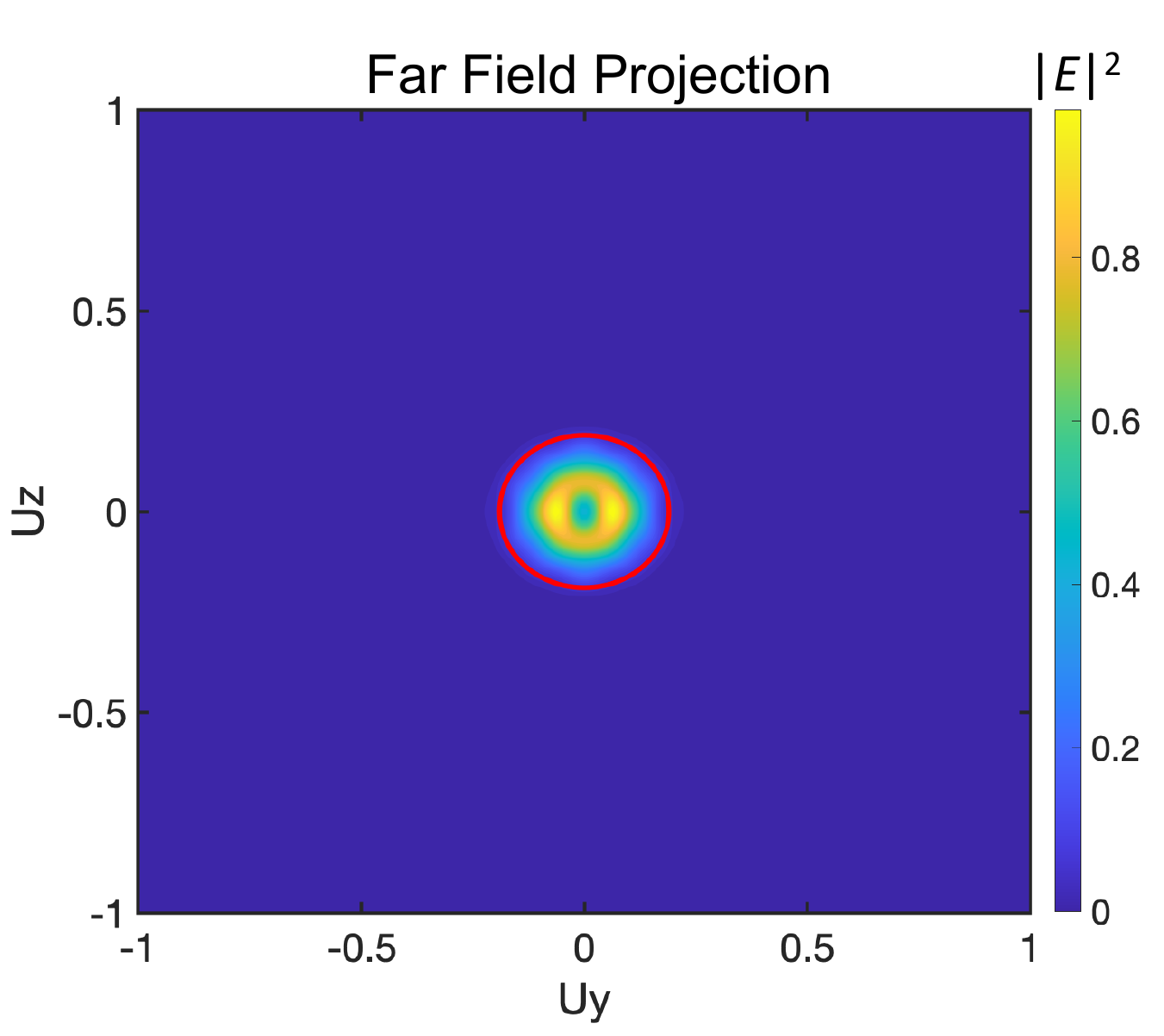}
    \caption{\label{fig:wide}Diamond waveguide far field projection simulations. $U_y$ and $U_z$ are the directional unit vectors and $|E|^2$ is normalized to 1. We find that the projected mode from our waveguides is contained in an NA = 0.26, denoted by the red circle in the plot.}
    \label{figS10}
\end{figure*}

We used finite element simulations (COMSOL Multiphysics) to determine the far field NA of our diamond waveguides for a measure of the scalability of our system, and find that our waveguides have a calculated NA = 0.26, as shown in Fig. \ref{figS10}.

\section{Improved grating coupler efficiencies}
\begin{figure*}
    \centering
    \includegraphics[width=0.5\textwidth]{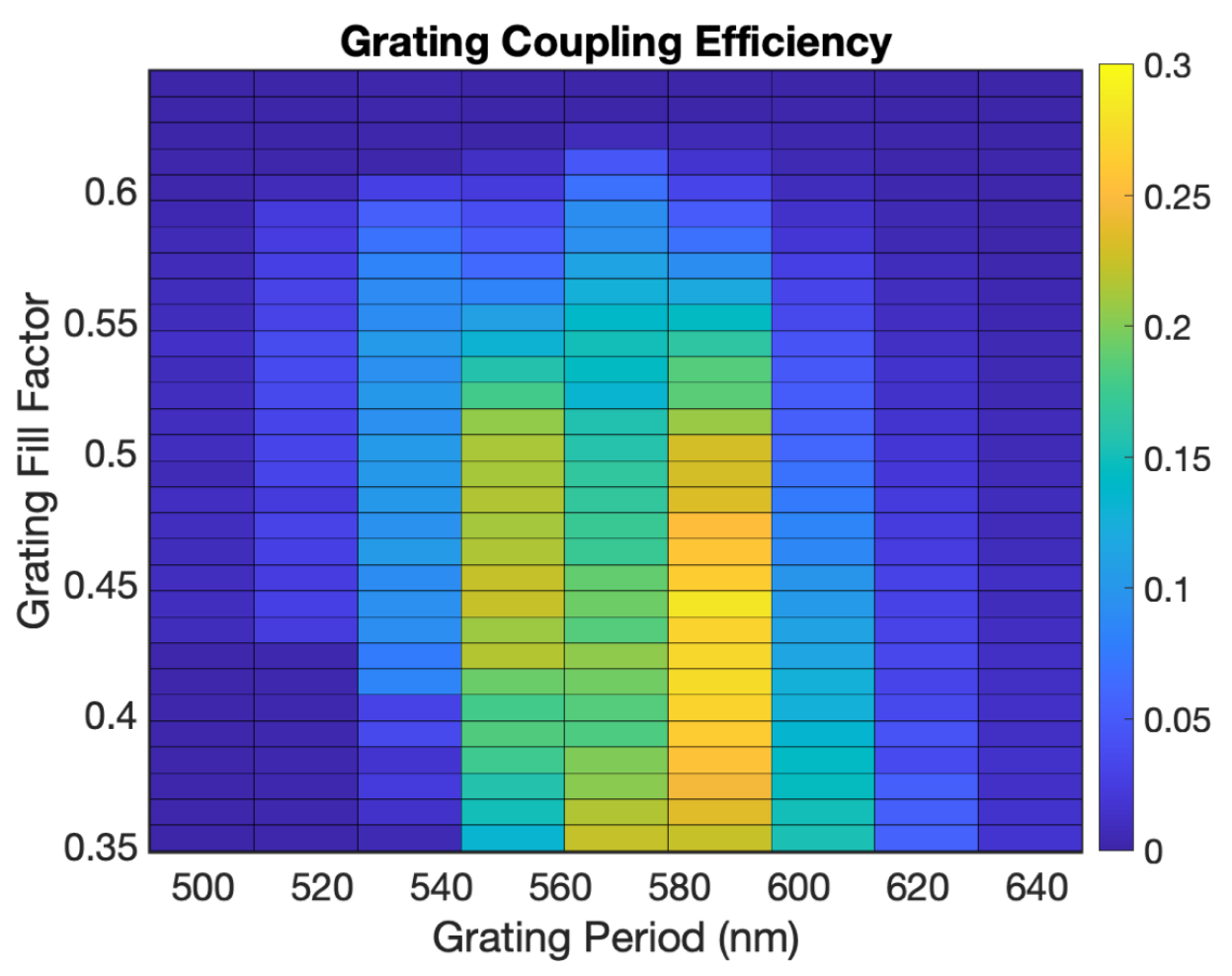}
    \caption{\label{fig:wide}Grating coupler optimization. The grating period and fill factor were swept around the simulated maximum value to find the experimentally best grating coupler. The highest measured coupling is 28.3$\%$ coupling at a period of 578 nm and a fill factor of 0.437.}
    \label{figS11}
\end{figure*}

To improve the insertion loss of our device, we ran a sweep of the period and fill factors of our grating couplers and measured the coupling efficiency. We measured the efficiency using loopback structures with the inputs and outputs measured through a fiber array. The results of this sweep are shown in Fig. \ref{figS11}. The highest measured coupling is 28.3$\%$ coupling at a period of 578 nm and a fill factor of 0.437. To improve this coupling even further, we experimented with depositing an index matching fluid (n = 1.52) on the grating couplers, thick enough that the fiber tips are completely submerged in the fluids when aligned for coupling. Using this technique, we are able to measure 41$\%$ grating coupler efficiency for grating couplers with a period of 621 nm and fill factor of 0.483.

\bibliography{SI}